\documentclass{PoS}

\PoS{PoS(LAT2005)151}

\usepackage{graphicx}

\title{Phase structure of lattice QCD with Wilson and Neuberger quarks at finite temperature and density
\thanks{This work is supported by the Key Project of National Science
Foundation (10235040), Key Project of National Ministry of
Education (105135), Project of the Chinese Academy of Sciences
(KJCX2-SW-N10) and Guangdong Natural Science Foundation
(05101821).} }

\ShortTitle{Phase structure of lattice QCD with Wilson and
Neuberger quarks ...}

\author{\speaker{Xiang-Qian Luo}\\
        Department of Physics, Zhongshan University, Guangzhou 510275, China\\
        E-mail: \email{stslxq@zsu.edu.cn}}

\author{He-Sheng Chen\\
        Department of Physics, Zhongshan University, Guangzhou 510275, China\\
Department of Physics, Yangzhou University, Yangzhou 225009,
China}

\author{Xiao-Lu Yu\\
        Department of Physics, Zhongshan University, Guangzhou 510275, China}

\abstract{We review our results for lattice QCD at finite
temperature and density from analytical and numerical calculations
with Wilson fermions and overlap fermions.}

\FullConference{XXIIIrd International Symposium on Lattice Field Theory\\
25-30 July 2005\\
Trinity College, Dublin, Ireland}

\begin{document}

\section{Introduction}

To study the nature of matter in QCD at high temperature $T$ or
large quark chemical potential $\mu$ is one of the most
challenging issues in particle physics. Several novel phases have
been proposed, such as quark-gluon plasma and color
superconductivity. Precise determination of the QCD phase diagram
on the ($\mu,T$) plane will provide valuable information for
experimental search.

In the continuum, the grand canonical partition function of QCD at
finite $T$ and $\mu$ is
\begin{eqnarray}
Z={\rm Tr} ~{\rm exp}\left(-  {H - \mu N \over T} \right),
\label{sixth}
\end{eqnarray}
where $N$ is particle number operator $N=\int d^3x ~
\psi^{\dagger}(x) \psi (x)$.

Monte Carlo (MC) simulation of lattice gauge theory (LGT) is the
most popular nonperturbative method based on the first principles.
This approach has been successfully applied to QCD at finite $T$
with zero $\mu$. However, LGT experiences serious problems, like
species doubling with naive fermions and complex action at finite
$\mu$.

The Hamiltonian formulation of LGT at finite $\mu$
 does not
have the complex action
problem\cite{Gregory:1999pm,Luo:2000xi,Fang:2002rk,Luo:2004mc}.
The complex action problem in Lagrangian formulation forbids
numerical simulation at real $\mu$. The recent years have seen
enormous
efforts\cite{Fodor:2001au,deForcrand:2002ci,D'Elia:2002gd,Chen:2004tb}
on solving the complex action problem.

There have been several popular approaches to solving the species
doubling problem of naive fermions. The staggered (KS) fermion
approach preserves the remnant of chiral symmetry, but it breaks
the flavor symmetry and doesn't completely solve the species
doubling problem. The Wilson fermion approach avoids the doublers
and preserve the flavor symmetry, but it explicitly breaks the
chiral symmetry; In order to define the chiral limit, one has to
do nonperturbative fine-tuning of the bare fermion mass.

The overlap fermion approach\cite{Narayanan,Neuberger} is claimed
to have the properties that chiral symmetry is preserved and
species doubling problem may be solved. However, the Dirac
operator is nonlocal, and the computational costs for simulating
dynamical overlap fermions are typically two orders of magnitude
heavier than for the Wilson or KS formulations. It is also very
tough to introduce the chemical potential into the action. Before
the breakthrough of numerical algorithms for applying overlap
fermions to QCD thermodynamics, it is very useful to do an
analytical study.

In this paper, we summarize our study on above issues using
Hamiltonian LGT with Wilson fermions, and Lagrangian MC
simulations with four flavors of Wilson fermions. We also present
some new results from strong coupling analysis of Lagrangian LGT
with overlap fermions.

\section{Hamiltonian lattice QCD with Wilson fermions}
\label{free_fermion_mu}

We begin with the QCD Hamiltonian with Wilson fermions $H=
H_g+H_f$ at $\mu=0$ on 1 dimensional continuum time and $d=3$
dimensional spatial discretized lattice, where
\begin{eqnarray}
H_g &=& {g^{2} \over 2a} \sum_{x}\sum_{j=1}^{d}\sum_{\alpha=1}^8
E^{\alpha}_{j}(x)E^{\alpha}_{j}(x) -{1 \over ag^{2}} \sum_{p} {\rm
Tr} \left(U_{p}+U_{p}^{+}-2\right),
\nonumber \\
H_f &=& {1 \over 2a} \sum_{x}\sum_{j=1}^{d}\bigg[\bar{\psi}(x)
\left(\gamma_j-r\right)U_j(x)\psi(x+{\hat j}) - \bar{\psi}(x)
\left(\gamma_j+r\right)U_{j}^{\dagger}(x-\hat{j})\psi(x-{\hat
j})\bigg]
\nonumber \\
&&+\left(m+ {rd \over a}\right) \sum_{x} \bar{\psi}(x)\psi(x) .
\label{first}
\end{eqnarray}

According to Eq. (\ref{sixth}), the role of the Hamiltonian at
finite $\mu$ is played by $H_{\mu}=H- \mu N$. Denoting $N_f$,
$N_c$, $V$ the number of flavors, colors, and spatial lattice
sites, and $\vert\Omega \rangle$ the vacuum state when $\langle
\Omega \vert H_{\mu} \vert \Omega \rangle$ is minimized. For free
massless Wilson fermions at $T=0$, we
obtained\cite{Gregory:1999pm} the energy $\langle \Omega \vert H
\vert \Omega \rangle = 2N_c N_f \sum_{p}\vert p \vert \left[
\Theta \left(\mu -\vert p \vert \right)-1\right]$, and the
subtracted energy density in the infinite volume limit $V\to
\infty$ and continuum limit $a\to 0$
\begin{equation}
\epsilon_{sub} = \frac{ \langle \Omega \vert H \vert \Omega
\rangle - \langle \Omega \vert H \vert \Omega \rangle
\vert_{\mu=0} } { N_c N_f V } = {2\over \left(2 \pi \right)^3}
\int ~ \vert p \vert ~ \Theta \left(\mu -\vert p \vert \right)~
d^3{\vec p} ={8 \pi \over \left(2 \pi \right)^3} \int_{0}^{\mu} ~
p^3 ~ dp ={\mu^4 \over 4 \pi^2},
\end{equation}
which agrees with the continuum theory. Therefore in the lattice
Hamiltonian formulation, the chemical potential could be
introduced in a natural way as in the continuum.

\begin{figure}[htb]
\begin{center}
\begingroup%
  \makeatletter%
  \newcommand{\GNUPLOTspecial}{%
    \@sanitize\catcode`\%=14\relax\special}%
  \setlength{\unitlength}{0.1bp}%
\begin{picture}(2339,1943)(0,0)%
\special{psfile=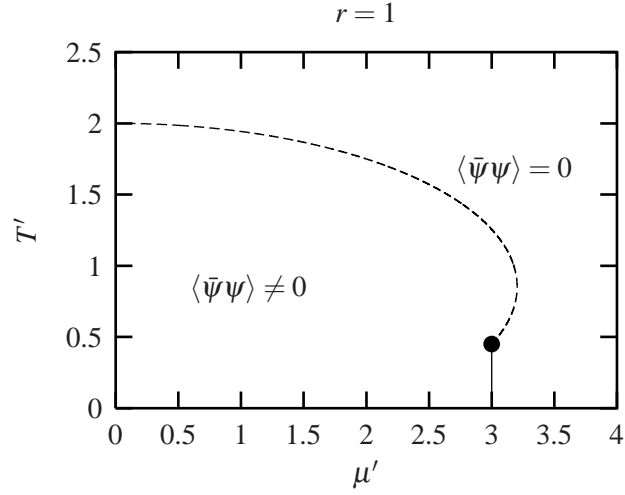 llx=0 lly=0 urx=468 ury=454 rwi=4680}
\put(1344,1793){\makebox(0,0){$r=1$}}%
\put(1344,50){\makebox(0,0){$\mu'$}}%
\put(900,750){\makebox(0,0){$\langle {\bar \psi} \psi \rangle \ne 0$}}
\put(1900,1200){\makebox(0,0){$\langle {\bar \psi} \psi \rangle = 0$}}
\put(100,971){%
\special{ps: gsave currentpoint currentpoint translate
270 rotate neg exch neg exch translate}%
\makebox(0,0)[b]{\shortstack{$T'$}}%
\special{ps: currentpoint grestore moveto}%
}%
\put(2289,200){\makebox(0,0){4}}%
\put(2053,200){\makebox(0,0){3.5}}%
\put(1817,200){\makebox(0,0){3}}%
\put(1581,200){\makebox(0,0){2.5}}%
\put(1345,200){\makebox(0,0){2}}%
\put(1108,200){\makebox(0,0){1.5}}%
\put(872,200){\makebox(0,0){1}}%
\put(636,200){\makebox(0,0){0.5}}%
\put(400,200){\makebox(0,0){0}}%
\put(350,1643){\makebox(0,0)[r]{2.5}}%
\put(350,1374){\makebox(0,0)[r]{2}}%
\put(350,1106){\makebox(0,0)[r]{1.5}}%
\put(350,837){\makebox(0,0)[r]{1}}%
\put(350,569){\makebox(0,0)[r]{0.5}}%
\put(350,300){\makebox(0,0)[r]{0}}%
\end{picture}%
\endgroup
 
\end{center}
\vspace{-0.5cm} \caption{Phase diagram from Hamiltonian lattice
QCD with massless Wilson fermions at strong coupling. The solid
and dotted lines stand respectively for the first and second order
transitions. The circle is the tricritical point.} \label{fig1}
\end{figure}

For infinitely strongly interacting Wilson fermions, integrating
out the gauge fields leads to four fermion
interactions\cite{Fang:2002rk}. Extreme conditions (large $T$ or
$\mu$) induce chiral phase transitions. For $N_f/N_c<1$ with
$N_c=3$, we obtained an equation for the critical line where the
chiral condensate and the dynamical mass of quark vanish
continuously\cite{Luo:2004mc}
\begin{eqnarray}
\mu_C^{\prime}= \left( 1+r^2 \right) \sqrt{
1-{\frac{2T_C^{\prime}}{1+3r^2} }} +T_C^{\prime}\ln {\frac{
1+\sqrt{ 1-{\frac{2T_C^{\prime}}{1+3r^2}}} }{1- \sqrt{
1-{\frac{2T_C^{\prime}}{1+3r^2}}}}} .
\label{second_order}
\end{eqnarray}
Here we have rescaled the chemical potential and temperature as
$\mu'=\mu /(3K /a)$ and $T'=T/(3K/a)$, with $K$ being the
effective coupling of four fermion interactions. Below some
$T^{\prime}_3$,  there is a first order chiral phase transition
line\cite{Luo:2004mc}
\begin{eqnarray}
\mu^{\prime}_C =1+2r^2 ,  \label{firstorder}
\end{eqnarray}
where the chiral condensate and the dynamical mass of quark vanish
discontinuously. The point when two lines described by Eq.
(\ref{second_order}) and Eq. (\ref{firstorder}) join at lower
$T^{\prime}$ is the tricritical point, as shown by the filled
circle in Fig. \ref{fig1}.

\section{Lagrangian lattice QCD with Wilson fermions}

The lattice action at $\mu=0$ proposed by
Wilson\cite{Wilson:1974sk} is $S=S_g+S_f$, where
\begin{eqnarray}
S_g &=& -{\beta \over 6} \sum_p {\rm Tr} (U_{p}
+U_{p}^{\dagger}-2),
\nonumber\\
S_f &=& \sum_{x,y} {\bar \psi}(x) M_{x,y}\psi(y),
\nonumber\\
    M_{x,y}&=&\delta_{x,y} -
\kappa \sum_{j=1}^{4} \bigg[
(r-\gamma_{j})U_{j}(x)\delta_{x,y-\hat{j}}  +
(r+\gamma_{j})U_{j}^{\dagger}(x-\hat{j})\delta_{x,y+\hat{j}}
\bigg] ,
\label{quark}
\end{eqnarray}
with $\beta=6/g^2$ and $\kappa=1/(2ma+8r)$. The lattice
Hamiltonian $H$ in Eq. (\ref{first}) could also be derived from
the Wilson action by Legendge transformation.

However, naive introduction of the chemical potential would lead
to divergent $\epsilon_{sub}$ in the continuum limit. In Ref.
\cite{Hasenfratz:1983ba}, the chemical potential is introduced by
replacing the link variables in the temporal direction in fermion
action in Eq. (\ref{quark}) with $U_4(x)\rightarrow e^{a\mu}
U_4(x)$ and $U^{\dag}_4(x)\rightarrow e^{-a\mu} U^{\dag}_4(x)$.
The fermionic action is reduced to the continuum one when $a \to
0$.

Nevertheless, the effective fermionic action in the partition
function becomes complex, and forbids MC simulation with
importance sampling. Several revised methods, e.g., improved
reweighting \cite{Fodor:2001au} and
  imaginary chemical potential methods\cite{deForcrand:2002ci,D'Elia:2002gd},
  were proposed to simulate QCD with KS fermions at
finite $\mu$.

Lattice QCD at imaginary chemical potential $i\mu_I$ does not
suffer the complex action problem. In Ref. \cite{Chen:2004tb}, we
applied this method to the MC study of  the phase structure of
$N_f=4$ Wilson fermions. We measured the expectation of the
Polyakov loop, chiral condensate and their susceptibilities for
various $(\mu_I,T)$ at some $\kappa$. From the position of the
peak in the susceptibilities, we determine the transition point.
Replacing $\mu_I$ by $-i\mu$, we directly continue the transition
line on the $(\mu_I,T)$ plane to the real $(\mu,T)$ plane.

\begin{figure} [htbp]
\begin{center}
\includegraphics[totalheight=2.0in]{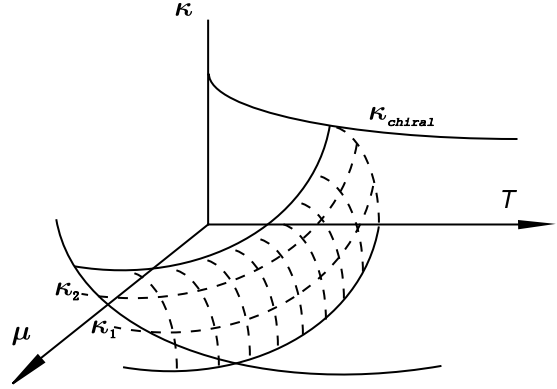}
\end{center}
 \caption{Phase diagram of lattice QCD with
four flavors of Wilson quarks in the $(\mu, T, \kappa)$ parameter
space, from MC simulations. For $\kappa \in [0,\kappa_1]$ and
$\kappa \in [\kappa_2,\kappa_{chiral}]$, the phase transition is
of first order, and while for $\kappa \in (\kappa_1,\kappa_2)$,
the transition is a crossover.} \label{fig21}
\end{figure}

Figure \ref{fig21} is the expected phase diagram of lattice QCD
with Wilson fermions in  the $(\mu, T, \kappa)$ parameter space.
There is a surface $\kappa=\kappa_{chiral}$ where the pion becomes
massless. Above this surface, there is no phase transition, as
confirmed by our numerical simulations for $\kappa=0.25$.
Interesting physics is below this surface. Of course, the order of
transition depends on the value of $\kappa$. For $r=1$, we find
$\kappa_1 \in (0.001, 0.15)$, $\kappa_2 \in (0.15, 0.165)$ and
$\kappa_{chiral} \in (0.17, 0.25)$.

\section{Lagrangian lattice QCD with overlap fermions}
\label{sec2}

The overlap fermionic action $S_f$ has the form\cite{Neuberger}
\begin{eqnarray}
S_f&=&
m\sum{\bar\psi}(x)\psi(x)+\sum_{x,y}{\bar\psi}(x)D(x,y)\psi(y),
\nonumber \\
D&=&1+X\frac{1}{\sqrt{X^{\dag}X}},
 \label{eqn1}
\end{eqnarray}
where $a$ is set to be 1 for convenience. The operator $D$ is
nonlocal and it is extremely difficult to do analytical
calculations. In Ref. \cite{Yu:2005eu}, we used the Taylor
expansion trick\cite{Ichinose,Nagao} to derive an overlap action
at finite $\mu$
\begin{eqnarray}
S_f&=&\left(1+\frac{m}{2}\right)\frac{C}{\vert A\vert}
\bigg(\sum_{x}\sum_{j=1}^{d}[{\bar q}(x)\gamma_{j}U_{j}(x)
q(x+\widehat{j}) -{\bar
q}(x+\widehat{j})\gamma_{j}U_{j}^{\dag}(x)q(x)]
 \nonumber\\
 & & ~~~~~~~+\sum_{x}[e^{\mu}{\bar q}(x)\gamma_{4}U_{4}(x)q(x+\widehat{4})
 -e^{-\mu}{\bar q}(x+\widehat{4})\gamma_{4}U_{4}^{\dag}(x)q(x)]
\bigg)+m\sum_{x}{\bar q}(x)q(x),
 \label{eqn16}
\end{eqnarray}
where $A=4r-M_{0}$, and $C=t/2$, with $t$ an expansion parameter.
The fermion fields $q$ and ${\bar q}$ are related to $\psi$ and
${\bar \psi}$ by \cite{Nagao}
\begin{eqnarray}
\bar{q}={\bar \psi},  ~~ q=\left(1-\frac{1}{2}D\right)\psi .
\label{eqn12}
\end{eqnarray}
The chiral order parameter is then given by
\begin{eqnarray}
\langle \bar{q}q \rangle =\langle {\bar
\psi}\left(1-\frac{1}{2}D\right)\psi \rangle  . \label{eqn14}
\end{eqnarray}

\begin{figure} [htbp]
\begin{center}
\includegraphics[totalheight=2.6in]{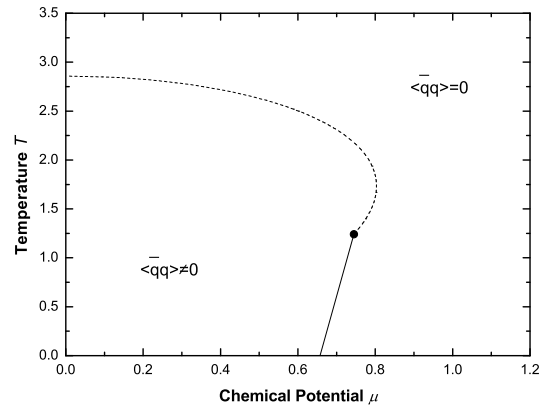}
\end{center}
\vspace{-1cm} \caption{Phase diagram of lattice QCD with massless
overlap fermions at strong coupling on the $(\mu,T)$ plane. The
dotted and solid lines stand respectively for the second and first
order transitions. The circle is the tricritical point.}
\label{fig3}
\end{figure}

In Ref. \cite{Yu:2005eu}, we studied the phase structure of LGT
with overlap fermions on the $(\mu,T)$ plane at the strong
coupling.
 The phase diagram is shown in Fig.
\ref{fig3}. We find that the phase structure is very similar to
the Wilson fermion case.

\section{Discussions}

LGT with imaginary chemical potential could be used for simulating
QCD at small $\mu$. The overlap fermion approach is a promising
one for investigating chiral properties of the phase transition,
and the Hamiltonian lattice formulation is a more natural way to
introduce chemical potential; However, a lot of efforts have to be
made before realistic MC simulations could be carried out.

\end{document}